\documentclass[12pt]{article}
\usepackage[left=2.5cm,top=2.50cm,right=2.5cm,bottom=2.50cm]{geometry}
\usepackage{mathrsfs}
\usepackage{amsmath,amssymb,latexsym,color,cancel,graphicx,bbm,colortbl}
\usepackage[english]{babel}
\usepackage[latin1]{inputenc}
\usepackage{ragged2e}
\usepackage{cite}
\usepackage{subfig}
\begin{document}
\date{}

\title{The Dunkl-Fokker-Planck Equation in $1+1$ Dimensions}
\author{R. D. Mota$^{a}$,   D. Ojeda-Guill\'en$^{b}$\footnote{{\it E-mail address:} dojedag@ipn.mx}, and M. A. Xicot\'encatl $^{c}$} \maketitle

\begin{minipage}{0.9\textwidth}
\small $^{a}$ Escuela Superior de Ingenier{\'i}a Mec\'anica y El\'ectrica, Unidad Culhuac\'an,
Instituto Polit\'ecnico Nacional, Av. Santa Ana No. 1000, Col. San
Francisco Culhuac\'an, Alc. Coyoac\'an, C.P. 04430, Ciudad de M\'exico, Mexico.\\

\small $^{b}$ Escuela Superior de C\'omputo, Instituto Polit\'ecnico Nacional,
Av. Juan de Dios B\'atiz esq. Av. Miguel Oth\'on de Mendiz\'abal, Col. Lindavista,
Alc. Gustavo A. Madero, C.P. 07738, Ciudad de M\'exico, Mexico.\\

\small $^{c}$ Departamento de Matem\'aticas, Centro de Investigaci\'on y  Estudios Avanzados del IPN, C.P. 07360, Ciudad de M\'exico, Mexico.\\

\end{minipage}

\begin{abstract}
By replacing the spatial derivative with the Dunkl derivative, we generalize the Fokker-Planck equation in (1+1) dimensions.
We obtain the Dunkl-Fokker-Planck eigenvalues equation and solve it for the harmonic oscillator plus a centrifugal-type potential. Furthermore, when the drift function is odd, we reduce our results to those of the recently developed Wigner-Dunkl supersymmetry.
\end{abstract}


\section{Introduction}
Many physical and biological systems can be modeled by stochastic differential equations. Among them is the Fokker-Planck differential equation which describes the transition probability function of the stochastic process. The Brownian motion has been the main physical problem described by the Fokker-Planck equation (FPE), or one of its variant forms\cite{libro1,libro2,libro3,risken,junker}. Several methods are used to find solutions for the FPE:  analytical methods, transformation of the FPE to a Schr\"odinger-type equation and numerical methods \cite{risken}. When the FPE can be transformed to a Schr\"odinger-type equation,  group theory \cite{gt1,gt2}, supersymmetric quantum mechanics and shape invariance \cite{junker,prl,polloto,anjos} are quantum mechanical methods that have been used successfully to solve the FPE for the so called confined potentials \cite{risken}.

On the other hand, in recent years theoretical research has been done in quantum mechanics by replacing the partial derivative $D_x\equiv\frac{\partial}{\partial x}$ by the Dunkl derivative $D_x\equiv\frac{\partial}{\partial x}+\frac{\mu}{x}(1-R)$  \cite{GEN1,GEN2,GEN3,GEN4,GAZ1,GAZ2,GAZ3,SCH,SCH2}, where $R$ is the reflection operator whose action on any function is given by $f(x)$ is $Rf(x)=f(-x)$. Even more, the Dunkl derivative has been applied to study statistical and thermodynamic properties of physical problems, as can be seen in Refs. \cite{DONGT,HAM1,HAM2,BER,QUES,QUES2,JUNK,HAM3,HAM4,HAM5,HAM6}. The motivation for these works is that the parameter introduced by the Dunkl derivative may be helpful to adjust the theoretical results to better match the experimental results.

The Fokker-Planck equation has been used to study systems of a few particles. In particular, in Ref. \cite{carmichel} the Fokker-Planck equation has been applied to systems of about $N=300$ particles, which is small enough for quantum mechanical effects to appear, but not so small to be able to apply the Fokker-Planck equation to their study. Also, the Fokker-Planck equation has been applied to study the generalized Jaynes-Cummings model (a two-level atom interacting with a quantized mode of an optical cavity or a bosonic field) \cite{wang,dalton} and to systems of $N$-atoms of two-levels interacting with the quantized radiation, known as the Tavis-Cummings model \cite{wang,larson}.

The study of the energy spectrum and eigenfunctions of few-body systems is usually obtained by developing their eigenfunctions in terms of a base of a single particle eigenfunctions \cite{mosh,garci}. For this reason it makes sense to obtain exact solutions of equations of a single particle, relativistic or non-relativistic, and in particular with the Dunkl derivative, as can be seen in Refs. \cite{GEN1,GEN2,GEN3,GEN4,GAZ1,GAZ2,GAZ3,SCH,SCH2,DONGT,HAM1,HAM2,BER,hamil,merad,schulze}. Due to the previous discussion, it is important to find
the Fokker-Planck equation eigenfunctions, which in itself is difficult to solve \cite{risken}. In this context, the objective of the present work is to obtain what we call the Dunkl-Fokker-Planck equation (DFPE) by changing the standard derivative in the Fokker-Planck equation by the Dunkl derivative. Then, we study the supersymmetry of the DFPE and solve it exactly for the generalized harmonic oscillator potential $V(x)=\frac {a( a-1) }{x^2}+x^2-2a-1$.

The paper is organized as follows. In Sec. 2, we reduce the FPE to the standard form of the Schr\"odinger equation of supersymmetric quantum mechanics. We identify the drift potential with the superpotential of supersymmetric quantum mechanics. This identification is very important for the applications of the DFPE. In Section 3, by assuming an odd superpotential $w(x)$, we particularize our results of the DFPE. We show that in this particular case, our results match those obtained with supersymmetric Dunkl-Wigner quantum mechanics introduced in Ref. \cite{shi}. As an application of our results we solve the DFPE in a closed way for our generalized harmonic oscillator potential in Section 4. Finally, we give some concluding remarks in Sec. 5.

\section{The Dunkl-Fokker-Planck equation in $1+1$ dimensions}

In one dimension the Fokker-Planck equation for the probability density
\begin{equation}
\frac{\partial \mathcal{P}(x,t)}{\partial t} =\left(-\frac{\partial }{\partial x}D^{(1)}(x)+\frac{\partial^2}{\partial x^2} D^{(2)}(x)\right)\mathcal{P}(x,t), \label{FP}
\end{equation}
where $D^{(1)}(x)$ and $D^{(2)}(x)$ are the drift and diffusion functions, respectively. Since it is usual to set $D^{(2)}(x)=$constant \cite{polloto,anjos,englefield,ho}, we set $D^{(2)}(x)=1$ and the drift coefficient to be given as
\begin{equation}
 D^{(1)}(x)=2w(x) \label{driftsuper}.
\end{equation}
If we propose the probability density to be given by
\begin{equation}
\mathcal{P}(x,t)= e^{-\lambda t}e^{\int w(x)dx}\phi(x),
\end{equation}
we arrive to the eigenvalues  equation
\begin{equation}
H\phi(x)\equiv \left(-\frac{d^2 }{d x^2}+w(x)^2+w'(x)\right)\phi(x)=\lambda\phi(x),  \label{susy}
\end{equation}
which is formally equal to the Schr\"odinger equation written in terms of the supersymmetric quantum mechanics  superpotential $w(x)$. This fact has  allowed to use the theory and results of the standard supersymmetric quantum
mechanics to solve the Fokker-Planck equation, see for instance \cite{risken,prl,junker,englefield,nicola,ioffe}. It is immediate to show that $\phi_0=e^{\int w(x)dx}$ satisfies
$H\phi_0=0$, and therefore, $\phi_0$ satisfies equation (\ref{susy}) with eigenvalue $\lambda=0$. This version of the Fokker-Planck equation, although perhaps in another notation, is well known \cite{polloto,anjos,ho}. So we have just recalled it here in the notation of our paper.\\

The purpose of  the present paper is to study the Fokker-Planck equation and its solutions by changing the standard derivative  $\frac{\partial}{\partial x}$ by the Dunkl derivative
\begin{equation}
D_x\equiv\frac{\partial}{\partial x}+\frac{\mu}{x}(1-R), \label{dun}
\end{equation}
where the parameter $\mu$ satisfies $\mu>-1/2$ \cite{GEN4}, and $R$ is the reflection operator with respect to the $x$-coordinate.
The action of $R$ on any function $f(x)$ is $Rf(x)=f(-x)$, and therefore
\begin{equation}
 R^2=1, \hspace{3ex}\frac{\partial}{\partial x}R=-R\frac{\partial}{\partial x},\hspace{3ex}Rx=-xR, \hspace{3ex}R D_x=-D_xR. \label{pro1}
\end{equation}

 By a  direct calculation, we show that the second order Dunkl derivative is given by
\begin{equation}
D_x^2=\frac{\partial^2}{\partial x^2}+\frac{2\mu}{x}\frac{\partial}{\partial x}-\frac{\mu}{x^2}(1-R). \label{lap}
\end{equation}
The substitution of these expressions into equation (\ref{FP}) leads to obtain the time-dependent Dunkl-Fokker-Planck equation
\begin{equation}
\frac{\partial \mathcal{P}(x,t)}{\partial t} =\left(\frac{\partial^2}{\partial x^2}+\frac{2\mu}{x}\frac{\partial}{\partial x}-
\frac{\mu}{x^2}\left(1-R\right) -2\frac{\partial}{\partial x} w(x)-\frac{2\mu}{x}\left(1-R\right)w(x)\right)\mathcal{P}(x,t) \label{DFP}.
\end{equation}
Now, if we introduce in this equation the separable variables product for the density probability
\begin{equation}
\mathcal{P}(x,t)=e^{-\lambda t}\psi(x),\hspace{5ex} \lambda>0
\end{equation}
we obtain the Dunkl-Fokker-Planck eigenvalues equation
{\footnotesize
\begin{equation}
\left(-\frac{d^2}{d x^2}-\frac{2\mu}{x}\frac{d}{d x} + \frac{\mu}{x^2}(1-R)+2\left(\frac{d w(x) }{d x}\right)+2w(x)\frac{d}{d x}+\frac{2\mu w(x)}{x}-\frac{2\mu}{x}(R w(x))R\right)\psi(x)=\lambda\psi(x),\label{DFP2}
\end{equation}
}where the last term in the left hand side becomes from the fact
\begin{equation}
Rw(x)\psi(x)=w(-x)\psi(-x)=(Rw(x))(R\psi(x)).\label{parity}
\end{equation}
As it is expected, we recover the Fokker-Planck equation  (\ref{FP})  written in terms of the superpotential $w(x)$ when the Dunkl parameter $\mu$ vanishes. Also, since $w(x)=D^{(1)}(x)/2$ is the standard supersymmetric quantum mechanical superpotential, we can use this fact as a guide to search for drift functions which lead to the DFPE being exactly solvable.  An important point that must be emphasized is that to derive equation (\ref{DFP2}) we do not suppose any parity property on the superpotential function $w(x)$.

We can write equation (\ref{DFP2}) in the form $H_{DFP}\psi(x)=\lambda\psi(x)$. Since  we are interested in finding the definite parity eigensolutions for the DFP equation, the commutation relation $[P,H_{DFP}]=0$ must be hold. This condition imposes that the superpotential $w(x)$ must be set as an odd function.

In one-dimensional Dunkl quantum mechanics the scalar product  it is defined by \cite{GEN4,hassa}
\begin{equation}
<f|g>\int_{-\infty}^{\infty}f^*(x)g(x)|x|^{2\mu}dx. \label{norm}
\end{equation}
The functions $f(x)$ and $g(x)$ for which this integral is finite belong to the $\mathcal{L}^2$-space associated with this scalar product.  Therefore, the  DFPE eigenfunctions must be normalized according to
\begin{equation}
\int_{-\infty}^{\infty}\psi^*_\lambda(x)\psi_{\lambda'}(x)|x|^{2\mu}dx=\delta_ {\lambda\lambda'}.
\end{equation}
Notice that the scalar product (\ref{norm}) is such that the Dunkl linear momentum $iD_x$ is an Hermitian operator.

\section{The Dunkl-Fokker-Planck equation and Wigner-Dunkl supersymmetry}

In this Section we will obtain the supersymmetric quantum mechanics  analogue of the FPE (see equation (\ref{susy})). To do this, we strictly assume that the
superpotential $w(x)$ is an odd function and therefore $Rw(x)\psi(x)=w(-x)\psi(-x)=-w(x)R\psi(x)$. Thus, by setting
\begin{equation}
\psi(x)=e^{\int w(x) dx}\Psi(x),
\end{equation}
our Dunkl-Fokker-Planck equation, equation (\ref{DFP2}), takes the form \begin{equation}
H\psi(x)=\lambda \psi(x),
\end{equation}
where
\begin{equation}
H=-\frac{d^2}{d x^2}-\frac{2\mu}{x}\frac{d}{d x} + \frac{\mu}{x^2}(1-R)+ w(x)^2+w'(x)+\frac{2\mu}{x}w(x)R.\label{us}
\end{equation}
We performed this particularization to match our results with those of the Wigner-Dunkl supersymmetry, reported by Dong {\it et. al.} in Ref. \cite{shi}.

On the other hand, starting with the generalized Dunkl supersymmetric quantum mechanics operators
\begin{eqnarray}
&&A\equiv \frac{1}{\sqrt{2}}(D_x+w(x)),\\
&&A^\dagger \equiv \frac{1}{\sqrt{2}}(D_x-w(x)),
\end{eqnarray}
and assuming $w(x)$ is an odd function, the supersymmetric quantum mechanics partner Hamiltonians $H_+\equiv AA^\dagger$ and
$H_-\equiv A^\dagger A$ have been reported in Ref. \cite{shi}, and are given by
\begin{equation}
H_\pm=\frac{1}{2}\left[-\frac{d^2}{d x^2}-\frac{2\mu}{x}\frac{d}{d x} + \frac{\mu}{x^2}(1-R)+ w(x)^2\pm \left(w'(x)+\frac{2\mu}{x}w(x)R\right)\right].
\end{equation}
A simple comparison of these partner Hamiltonians with the DFP Hamiltonian-type equation (\ref{us}) leads to conclude that $H=2H_+$.
Thus, in principle, we can apply the Wigner-Dunkl supersymmetry and shape invariance results developed in Ref. \cite{shi}, to obtain the solutions for our Dunkl-Fokker-Planck equation (\ref{DFP2}).
However, the literature on supersymmetric quantum mechanics leads us to the conclusion that there are several superpotentials that do not have defined parity\cite{junker,cooper,bagchi}.
Moreover, it can be shown that the restriction that $w(x)$ is an odd function implies that the shape invariance conditions of Ref. \cite{shi} hold only for the standard harmonic oscillator. Therefore, the solution of the DFPE must be searched  by the different analytical and perturbative methods used to solve de Schr\"odinger equation.

\section{The DFPE for the harmonic oscillator plus a centrifugal-type potential}

In this Section we solve the Dunkl-Fokker-Planck  equation (\ref{DFP2}) in an analytical way. We find out exactly the eigenfunctions and the energy spectrum for a generalized  harmonic oscillator. To this end, we set the superpotential as $w(x)=\frac{a}{x}-x$. Then, the expression
\begin{equation}
w(x)^2+w'(x)=\frac {a( a-1) }{x^2}+x^2-2a-1,
\end{equation}
reproduces our generalized harmonic oscillator potential.\\

I) Even parity solutions.\\

These are obtained if we substitute $R\psi(x)=\psi(x)$ into equation (\ref{DFP2}), which simplifies to
\begin{equation}
- x^2\frac{d^2}{d x^2}\psi(x)+2x \left( -{x}^{2}+a-\mu \right) \frac{d}{d x}\psi(x) +4\left[  \left( -\mu-\frac{\lambda}{4}-\frac{1}{2}
 \right) {x}^{2}+a \left( \mu-\frac{1}{2} \right)  \right] \psi(x)=0.\label{so1}
\end{equation}
If we perform the variable change $u=x^2$, this equation transforms to
\begin{equation}
\frac{d^2}{d u^2}\psi(u)+\left(1 +\frac{2(\mu-a)+1}{2u} \right) \frac{d}{d u}\psi(u) +4\left(\frac{a(1-2\mu)}{2u^2}+\frac{4\mu+\lambda+2}{4u} \right) \psi(u)=0.
\end{equation}
By setting $\psi(u)=u^ae^{-u}g(u)$, we get the differential equation
\begin{equation}
u\frac{d^2}{d u^2}g(u)\left(a+\mu+\frac{1}{2}-u\right)\frac{d}{du}g(u)+\frac{\lambda}{4}g(u)=0.\label{ops}
\end{equation}
It is known that the differential equation
\begin{equation}
x \frac {{ d}^{2}}{{d}{x}^{2}}f\left( x \right)+ \left( \alpha+1 -x\right) {\frac
{ d}{{ d}x}}f\left( x \right) + n f\left( x \right) =0, \label{laguerre}
\end{equation}
has as solutions the Laguerre polynomials \cite{lebedev}
\begin{equation}
f(x) =  L_n^\alpha (x), \hspace{5ex}n=0,1,2,3...,\hspace{5ex}\alpha>-1.
\end{equation}
By direct comparison of equations (\ref{ops}) and (\ref{laguerre}) we find that
\begin{equation}
\alpha=a+\mu-\frac{1}{2},\hspace{8ex}\lambda=4n,\hspace{8ex}g(u)=L_n^\alpha(u). \label{esp}
\end{equation}
Thus, the solutions of equation (\ref{so1}) in the $x$-variable are given by
\begin{equation}
\psi(x)=C_e e^{-x^2}x^{2a}L_n^{a+\mu-1/2}(x^2), \label{solp}
\end{equation}
which is regular at the origin provided that $a>1/2$, $2a$ be an even integer and $a+\mu>-1/2$.\\

II) Odd parity solutions. \\

These are obtained if we substitute $R\psi(x)=-\psi(x)$ into equation (\ref{DFP2}), which takes the form
\begin{equation}
- x^2\frac{d^2}{d x^2}\psi(x)+2x \left( -{x}^{2}+a-\mu \right) \frac{d}{d x}\psi(x) -\left( (\lambda+2)x^2+2(a-\mu)\right) \psi(x)=0.\label{so2}
\end{equation}
Under the variable change $u=x^2$, we can write this equation as
\begin{equation}
\frac{d^2}{d u^2}\psi(u)+\left(1 +\frac{1-2(a-\mu)}{2u} \right) \frac{d}{d u}\psi(u) +\left(\frac{a-\mu}{2u^2}+\frac{\lambda+2}{4u} \right) \psi(u)=0.
\end{equation}
If we set $\psi(u)=u^{a-\mu}e^{-u}g(u)$, this differential equation reduces to
\begin{equation}
u\frac{d^2}{d u^2}g(u)\left(a-\mu+\frac{1}{2}-u\right)\frac{d}{du}g(u)+\frac{\lambda}{4}g(u)=0.\label{eps}
\end{equation}
Also, the comparison of equations (\ref{laguerre}) and (\ref{eps}) leads us to find
\begin{equation}
\alpha=a-\mu-\frac{1}{2},\hspace{8ex}\lambda=4n,\hspace{8ex}g(u)=L_n^\alpha(u). \label{esi}
\end{equation}
Hence, the solutions of equation (\ref{so2}) in terms of the $x$-variable are given by
\begin{equation}
\psi(x)=C_e e^{-x^2}x^{2(a-\mu)}L_n^{a-\mu-1/2}(x^2), \label{soli}
\end{equation}
which is regular at the origin provided that $a-\mu>1/2$, $2(a-\mu)$ be an odd integer and $a-\mu>-1/2$.

Therefore, although the DFPE is difficult to solve, we have found that the potential $V(x)=\frac {a( a-1) }{x^2}+x^2-2a-1$, which is obtained from the superpotential $w(x)=\frac{a}{x}-x$, is exactly solvable in terms of the Laguerre polynomials.

Let us note that to obtain the solutions (\ref{solp}) and (\ref{soli}), we have imposed explicitly that $R\psi(x)=+\psi(x)$ and $R\psi(x)=-\psi(x)$, respectively. However, for these solutions to have the required parity, some restrictions must be made on the parameters $a$ and $\mu$. For the solutions  (\ref{solp}) to be even for a given $\mu>-\frac{1}{2}$, we must impose that $a$=0,1,2,...,
and for the solutions  (\ref{soli}) to be odd, we must choose $a$ and $\mu$ in such a way that $2(a-\mu)=1,3,5,...$, which restricts the Dunkl parameter to take one of the values $\mu=a-\frac{1}{2}, a-\frac{3}{2}, a-\frac{5}{2},... >-\frac{1}{2}$.
In Fig. 1 we have plotted the even and odd functions, given by equations  (\ref{solp}) and (\ref{soli}) for $n$=0,1,2 with the choice $a$=1 and $\mu=\frac{1}{2}$.

\begin{figure}[ht]
 \centering
  \subfloat[]{
    \includegraphics[width=0.45\textwidth]{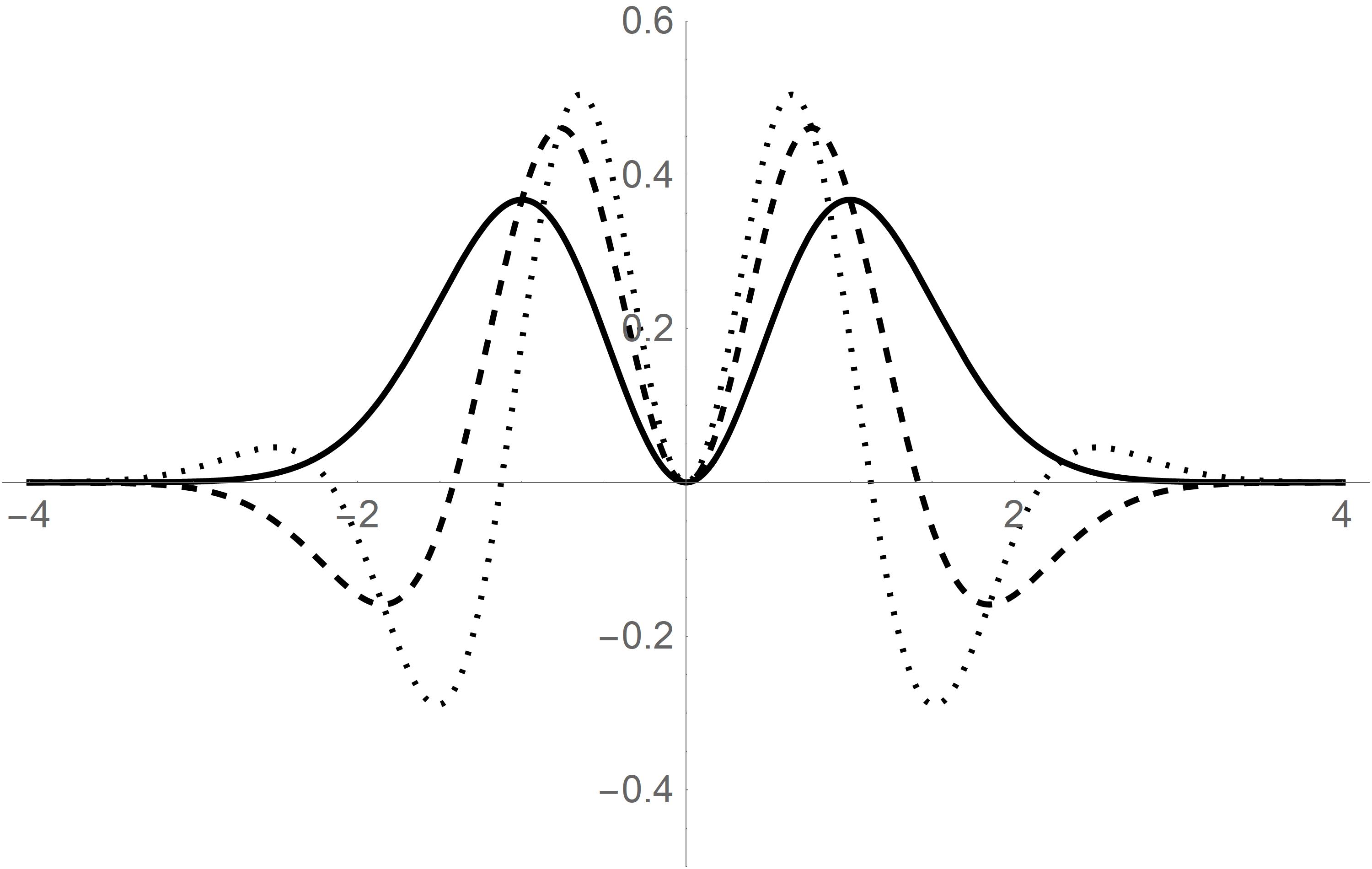}}
  \subfloat[]{
    \includegraphics[width=0.45\textwidth]{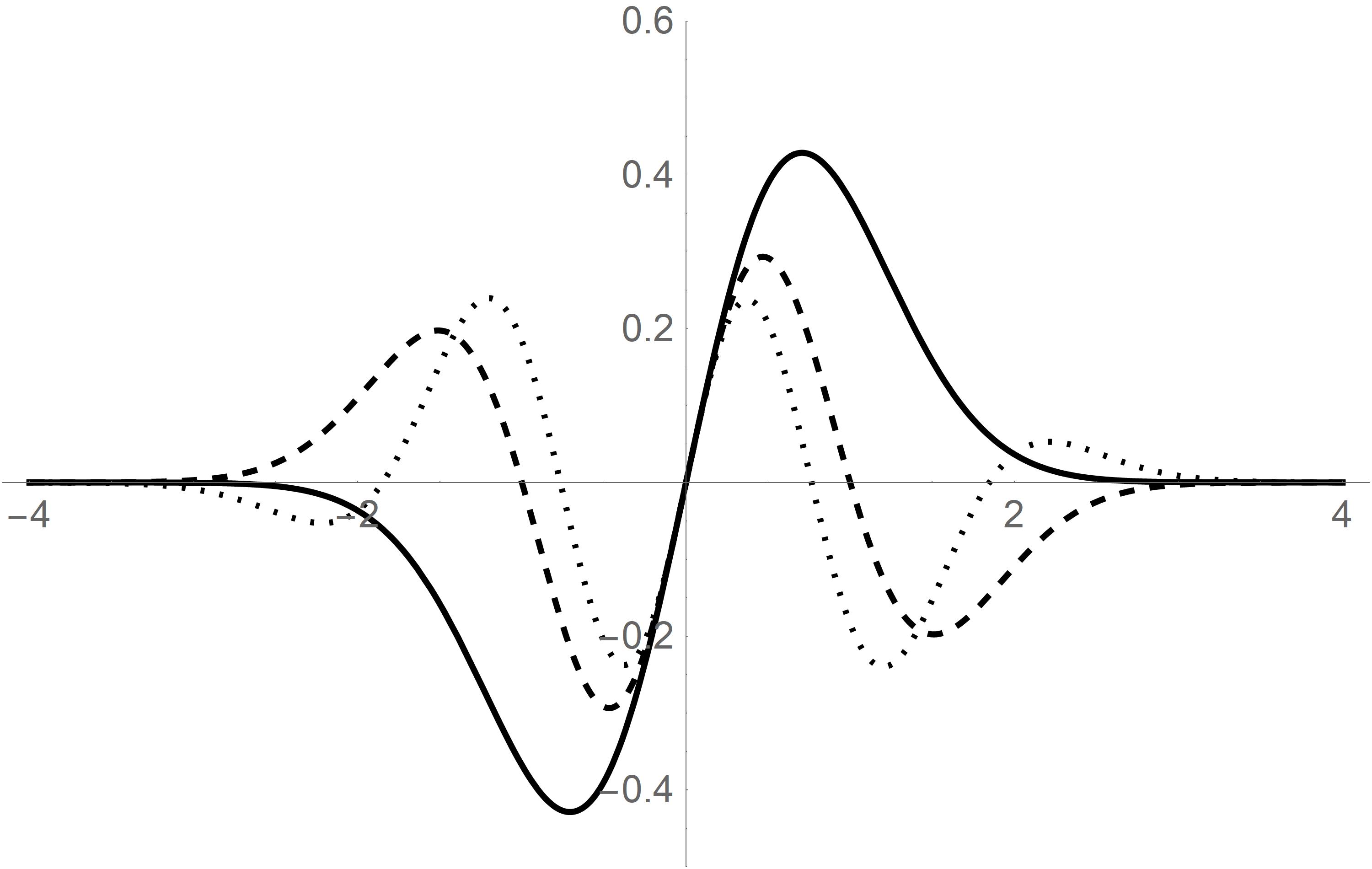}}
  \caption{\footnotesize First eigenfunctions of DFPE equation  with $a=1$, $\mu=\frac{1}{2}$. In Fig. (a) are plotted the even parity eigenfunctions for $n=0$ (solid), $n=1$ (dash), and $n=2$ (dot). In Fig. (b) are plotted the odd parity eigenfunctions $n=0$ (solid), $n=1$ (dash), and $n=2$ (dot).}
\end{figure}

Now, we shall obtain the solutions for the standard Fokker-Planck equation (\ref{FP}) with the superpotential $w(x)=\frac{a}{x}-x$. To this end, if we substitute $D^{(1)}(x)=2w(x)=2\left(\frac{a}{x}-x\right)$, $D^{(2)}(x)=1$ and $\mathcal{P}(x,t)\equiv e^{-\lambda t}\psi(x)$ into equation (\ref{FP}), we obtain
\begin{equation}
- x^2\frac{d^2}{d x^2}\psi(x)+2 \left(\frac{a}{x}-x\right) \frac{d}{d x}\psi(x) - 2\left(\frac{a}{x^2}-1\right)\psi(x)=\lambda \psi(x).
\end{equation}
With the definition  $\psi(x)= e^{-x^2}x^{2a}f(x)$, this equation transforms to
\begin{equation}
\frac{x}{2}\frac{d^2}{d x^2}f(x)+\left(a-x^2\right)\frac{d}{dx}f(x)+\frac{\lambda}{2}xf(x)=0.
\end{equation}
Finally, the change of variable $u=x^2$ lead us to the differential equation
\begin{equation}
u \frac { d^2}{d u^2}f\left( u\right)+ \left( a+\frac{1}{2} -u\right) \frac
{ d}{ d u}f\left( u\right) + \frac{\lambda}{4} f\left( u \right) =0.
\end{equation}
We can immediately identify this equation as that of the Laguerre polynomials (\ref{laguerre}), with
\begin{equation}
\alpha=a-\frac{1}{2},\hspace{10ex}\frac{\lambda}{4}=n. \label{esnd}
\end{equation}
Hence, the solutions of the standard Fokker-Planck equation for the superpotential $w(x)=\frac{a}{x}-x$ are given by
\begin{equation}
\psi(x)=C e^{-x^2}x^{2a}L_n^{a-1/2}(x^2)  \label{solnd}
\end{equation}
and the energy spectrum is given by
\begin{equation}
\lambda=4n.
\end{equation}
Thus, when we set $\mu=0$, equations (\ref{esp}), (\ref{solp}), (\ref{esi}) and (\ref{soli}) are appropriately reduce to those of equations (\ref{esnd}) and (\ref{solnd}). As far as we know, the solutions reported above for the standard Fokker-Planck equation (\ref{FP}) with the superpotential $w(x)=\frac{a}{x}-x$ have not been reported before.

\section{Concluding Remarks}
We have obtained a generalization of the Fokker-Planck equation in $1+1$ dimensions in terms of the Dunkl derivative. We have argued that setting the superpotential of the  standard supersymmetric quantum mechanics equal to the drift function, $w(x)=D^{(1)}(x)/2$, can  be used to search for suitable functions that exactly solve the DFPE. Assuming that the superpotential $w(x)$ is an odd function, we have shown that $H=2H_+$, being $H_+$ one of the partner Hamiltonians of the Wigner-Dunkl supersymmetry. Finally, the eigenfunctions and the energy spectrum of a generalized harmonic oscillator were found exactly.

The ideas developed in this work can be extended to apply the Dunkl-Fokker-Planck equation to study other physical systems  in (1+2) or (1+3) dimensions, or systems with space and time dependent drift and diffusion coefficients.
Furthermore, the Dunkl-Fokker-Planck equation introduced in this work can be studied in terms of more general derivatives, such as those reported in Refs. \cite{chung1,chung2}. Some of these ideas are works in progress and they will be submitted for review elsewhere.

\section*{Acknowledgments}
This work was partially supported by SNII-Mexico, COFAA-IPN, EDI-IPN, CGPI-IPN Project Numbers $20230633$, $20230732$, and CONAHCYT-Mexico grant CB-2017-2018-A1-S-30345.\\
We appreciate the observations made by the anonymous referee to improve our work.

\end{document}